\documentclass[a4,11pt]{article}
\usepackage{epsfig}
\usepackage{amssymb,epsfig,graphicx,epsf}

\usepackage{caption}
\usepackage{subcaption}
\usepackage{color}
\usepackage{amsfonts}
\usepackage{amsmath}
\usepackage{amssymb}
\usepackage{feynmf}
\newcommand{\be}{\begin{equation}}
\newcommand{\ee}{\end{equation}}

\newcommand{\ba}{\begin{eqnarray}}
\newcommand{\ea}{\end{eqnarray}}
\newcommand{\bal}{\begin{aligned}}
\newcommand{\eal}{\end{aligned}}

\begin{document}

\title{Local measurement of $\Lambda$ using pulsar timing arrays}

\author{Dom\`enec Espriu and Daniel Puigdom\`enech\\
Departament d'Estructura i Constituents de la Mat\`eria and \\ 
Institut de Ci\`encies del Cosmos (ICCUB)\\ Universitat de Barcelona\\
Mart\'\i ~i Franqu\`es, 1, 08028 Barcelona, Spain.}

\date{}

\maketitle

\begin{abstract}
We have considered the propagation of gravitational waves (GW) in de Sitter space-time
and how a non-zero value of the cosmological constant might affect their detection 
in pulsar timing arrays (PTA). If 
$\Lambda\neq 0$ waves are anharmonic in Friedmann-Robertson-
Walker coordinates and although the amount of this effect is very small it gives noticeable
effects for GW originating in extragalactic sources 
such as spiraling black hole binaries. The results indicate that the timing residuals induced 
by gravitational waves from such sources in PTA 
would show a peculiar angular dependence with a marked enhancement around a particular 
value of the angle subtended 
by the source and the pulsars, depending mainly on the actual value of the cosmological constant 
and the distance to the source. The position of the peak 
could represent a gauge of the value of $\Lambda$. The enhancement that the new effect brings 
about could facilitate the first direct detection of gravitational waves while representing 
a local measurement of $\Lambda$.
\end{abstract}

\vfill
\noindent
July 2012

\noindent
UB-ECM-FP-76/12

\noindent
ICCUB-12-317

\section{Introduction}
Pulsar timing arrays (PTA) are one of the most promising candidates to offer the first direct detection of gravitational 
waves. They have been collecting data already for almost a decade and they are expected to obtain signals in the next 
years. The idea behind PTA is to detect the correlated disruption of the periods measured 
for a significant number of pulsars due to the passing of a gravitational wave through the system 
\cite{hobbs1,hobbs2,hobbs3,lee}. The frequency range sensitive to this method is $10^{-9}s^{-1}\leq w\leq 10^{-7}s^{-1}$ \cite{hobbs1}, 
and the timing residual is expected to follow a power law \cite{hobbs2,jenet}. A key problem in making predictions 
for these signals is modeling in a realistic way the wave functions produced in the different sources, in particular 
the value of the amplitude of the metric perturbation $h$ is a free parameter in principle. Some bounds in the range of $10^{-17}\leq h\leq 10^{-15}$ have been set already \cite{jenet}. 

If $\Lambda\neq 0$ gravitational waves (GW) propagate in a de Sitter space-time not in flat Minkowskian space-time. 
The general practice is simply to account for the expansion of the universe by using a redshifted frequency 
according to the distance of the source \cite{lee}. In this work we go beyond this exceedingly simple approximation 
and use an approximate solution of the GW equation in de Sitter previously derived \cite{nos} and see that 
the conclusions change.

We assume that $\Lambda$ is somehow an intrinsic property of space-time rather than an effective description 
valid at extremely large scales. If so, it is expected to be present at virtually all scales, 
with the possible exception of 
gravitationally bound objects such as galaxies or local groups of galaxies. 
If $\Lambda$ is a fundamental constant of nature surely there should be a way
of determining locally its value. By `locally' here we mean at redshifts $z << 1$. This question has been
addressed in \cite{previous,ber} with varying conclusions. We will see that GW may open a nice window to realize this program. In fact, our 
results suggests that the currently observed non-zero value of $\Lambda$ may actually facilitate the first direct detection
of GW under some circumstances. 

This paper is organized as follows. In Section 2 the wave functions used are presented, the way in which 
the timing residuals are calculated is defined and a brief explanation of the coordinate systems involved is included. 
Section 3 is devoted to present our numerical analysis. In Section 4 we discuss the possibility of using 
this method to get some results on the value of the cosmological constant. Finally we sum up our conclusions 
in Section 5. 

\section{Gravitational waves and timing residuals with $\Lambda\neq 0$}
In Minkowski space-time, gravitational waves obey the simple wave equation $\Box h=0$. It is possible to show \cite{nos} 
that in de Sitter space-time with $\Lambda \neq 0$ and within the linearized approximation one can 
find solutions of the linearized Einstein equations in the traceless 
Lorenz gauge (TT gauge \cite{TT}) which obey the same equation of motion
\be\label{eom}
\Box h_{\mu\nu}^{SdS}=0.
\ee
Spherical massless waves are solution of this equation away from the source 
\be\label{gwdS}
h^{SdS}_{\mu\nu}=\frac{1}{r}\left(E_{\mu\nu}\cos[{w(t-r)}]+D_{\mu\nu}\sin[{w(t-r)}]\right).
\ee
However, as shown in 
\cite{nos}, this simple linearized solution only holds in a specific set of coordinates, the 
Schwarzschild-de Sitter (SdS) coordinates. This is easily seen by considering a linearized 
background solution (rather than wave-like solutions) 
and realizing that their unique static solution is the (linearized) 
Schwarzschild-de Sitter metric \cite{nos}.

Although providing a perfectly valid solution for gravitational waves, SdS coordinates are not adequate to 
make observational predictions. The proper isotropic and homogeneous coordinates are the Friedmann-Robertson-Walker (FRW) 
ones\footnote{Note that the FRW metric cannot be approximated to obey any linearized Einstein equation, see \cite{nos} for a detailed 
discussion.} and the solution (\ref{gwdS}) in such coordinates, neglecting $\mathcal{O}(\Lambda)$ and higher, reads
\be\label{gwfrw}
\bal
h^{FRW}_{\mu\nu}=&\frac{E_{\mu\nu}}{R}\left( 1+\sqrt{\frac{\Lambda}{3}}T\right)
\cos\left[{w(T-R)+w\sqrt{\frac{\Lambda}{3}}\left(\frac{R^2}{2}-TR\right)}\right] \\
& + \frac{D_{\mu\nu}}{R}\left( 1+\sqrt{\frac{\Lambda}{3}}T\right)
\sin\left[{w(T-R)+w\sqrt{\frac{\Lambda}{3}}\left(\frac{R^2}{2}-TR\right)}\right],
\eal
\ee
where $R$ is the usual radial FRW comoving coordinate and $T$ is cosmological time. 
Note that the linearization process that 
has been used makes sense as long as $\Lambda T^2,\ \Lambda R^2 << 1$ and also that in the TT gauge the 
only spatial components of the metric that are different from zero are the $X,Y$ entries
of the polarization tensors $E_{\mu\nu}$, $D_{\mu\nu}$. Although some temporal components 
of $E_{\mu\nu}$ and $D_{\mu\nu}$ are also non-zero in these coordinates, they are several orders of magnitude 
smaller than the spatial ones and therefore will not be relevant for the present study.

We note that the phase velocity of propagation of the GW in such coordinates is not $v_{p}=1$ but 
$v_{p}\sim 1-\sqrt{\frac{\Lambda}{3}}T + \mathcal{O}(\Lambda)$ \cite{nos}. On the other hand, with 
respect to the ruler distance travelled (computed with $g_{ij}$) the velocity is still $1$ (up to terms in $\Lambda$
of higher order to those considered).

\begin{figure}[htb]
\begin{center}
\includegraphics[scale=0.5]{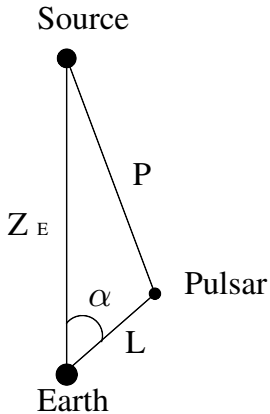}
\end{center}
\caption{Relative coordinates of the GW source (R=0), the Earth (located at $Z=Z_E$) with respect to the GW source
and the pulsar located at a coordinates $\vec P= (P_X, P_Y, P_Z)$ referred to the source. The $Z$ direction is chosen to be
defined by the source-Earth axis. Angles $\alpha$ and $\beta$ are the polar and azimuthal angles of the pulsar with 
respect to this axis.}
\label{coor} 
\end{figure}

Consider the set up depicted in Figure \ref{coor} describing the relative situation of a GW source 
(possibly a very massive black hole binary), the Earth and a nearby pulsar.
The timing residual \cite{tres} induced by (\ref{gwfrw}) will be given by
\be\label{tim}
H(T_E,L,\alpha,\beta,Z_E,w,\varepsilon,\Lambda)= -\frac{L}{2c} \hat n^i \hat n^j \int_{-1}^0\ dx\ h_{ij}^{FRW}(T_E +\frac{L}{c}x,
\vec P + L(1 + x)\hat n)
\ee
along the null geodesic from the pulsar to the earth, where we 
assume\footnote{This approximation is unessential and can be easily removed.} 
$\varepsilon \sim | E_{ij}|\sim | D_{ij}|$, $i,j=X,Y$ and the unit vector 
$\hat n$ is given by
$(-\sin\alpha \cos\beta,-\sin\alpha \sin\beta, \cos\alpha)$, $Z_{E}$ is the distance from the Earth to the source, $L$ the distance to the pulsar and $T_{E}$ the time of arrival of the wave to the local system. In deriving the previous timing residual we have neglected
the (non-zero) time components of $E_{\mu\nu}, D_{\mu\nu}$ that, as previously indicated, are several orders of magnitude
smaller.
The speed of light has been restored. We have 
assumed that from the pulsar to the Earth the electromagnetic signal follows the trajectory given by the line of sight 
$\vec R(x)= \vec{P}+L(1+x)\hat{n}$. Since we assume that within the Galaxy $\Lambda=0$, $L$ is also the ruler distance. 
Explicitly
\be\label{r}
\vec{R}(x)=\vec{P}+L(1+x)\hat{n}=(-x L \sin{\alpha}\cos{\beta},-x L \sin{\alpha}\sin{\beta},Z_{E}+x L \cos{\alpha})
\ee
or in modulus
\be\label{rmod}
R(x)=\sqrt{Z_{E}^2+2xLZ_{E}\cos{\alpha}+x^2L^2}\simeq Z_{E}+x L\cos{\alpha},
\ee
since we are considering $L<<Z_{E}$. This approximation does not affect in any significant way the results below.
We do not consider here the known contribution to the timing residual $H$ from the 
Earth's peculiar motion either. The integral is of course independent of the angle $\beta$ for any 
single pulsar but it will
depend on the relative angles when several pulsars are averaged. 

\begin{figure}[htb]
\centering
\includegraphics[scale=0.47]{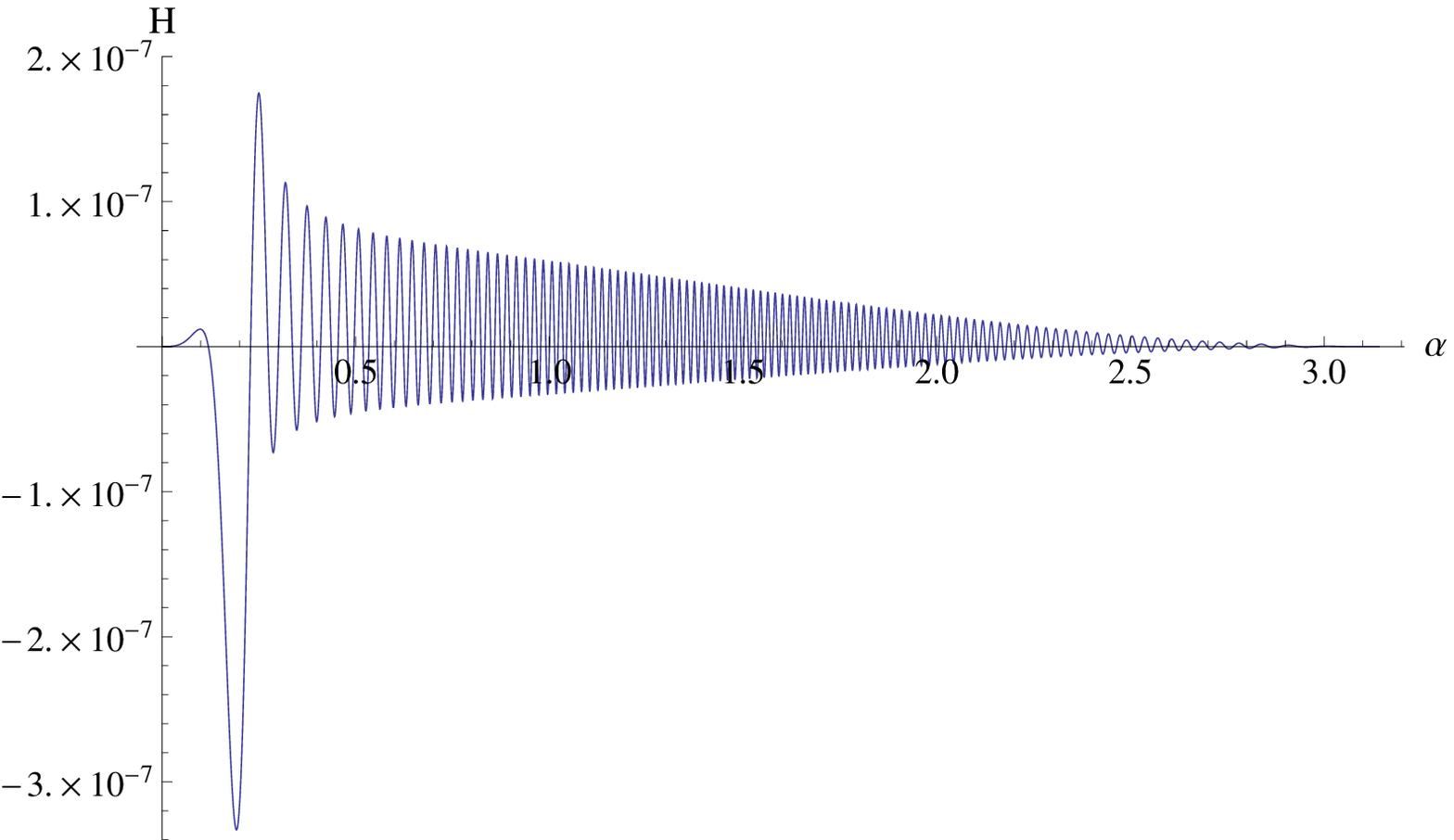}
\includegraphics[scale=0.47]{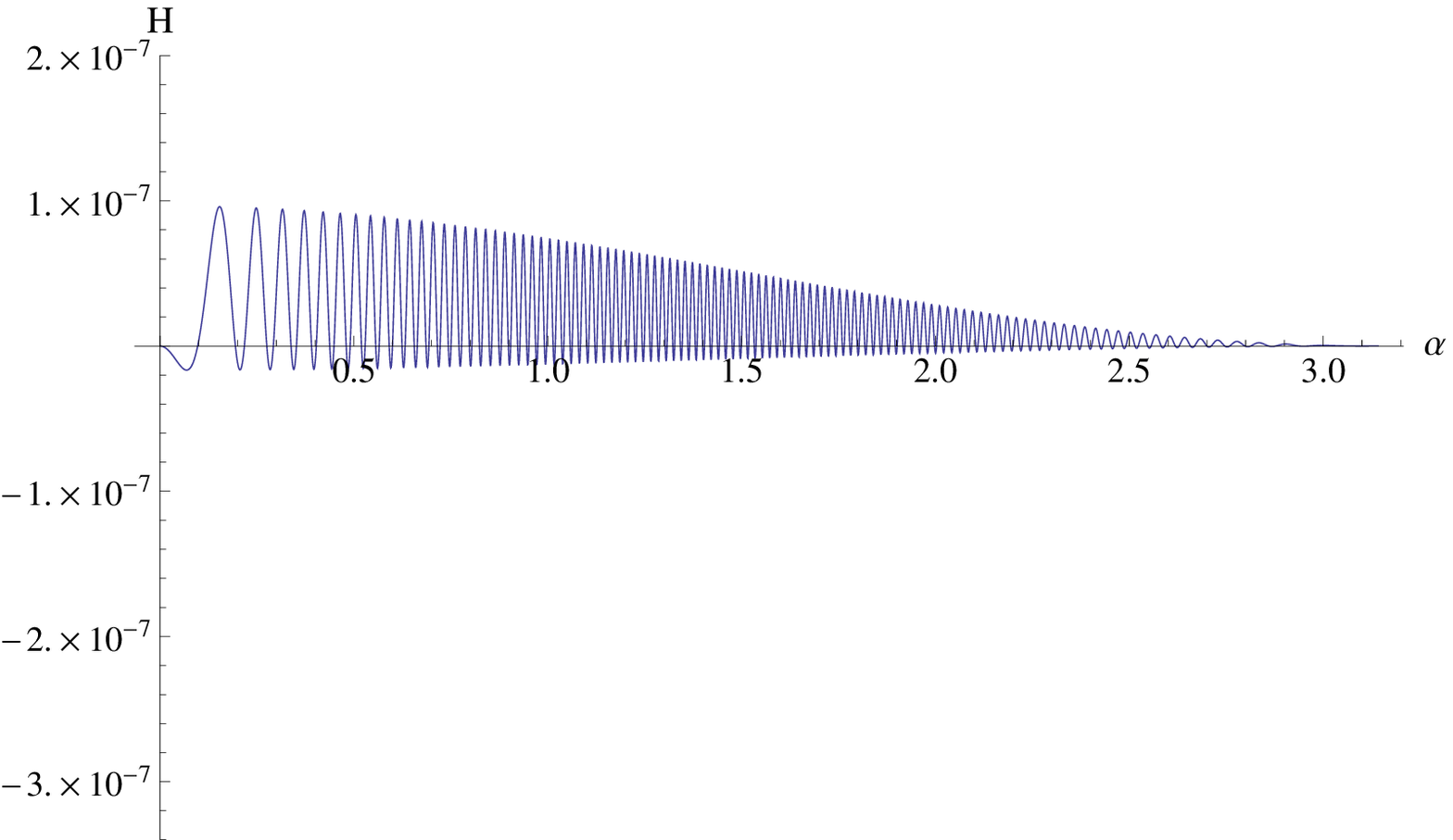}
\caption{On the left the raw timing residual for $\Lambda=10^{-35}s^{-2}$ as a function
of the angle $\alpha$ subtended by the source and the measured pulsar as seen from the observer. The figures are symmetrical for $\pi\leq\alpha\leq 2\pi$.
On the right the same timing 
residual for $\Lambda=0$. In both cases we take $\varepsilon=1.2\times 10^{9}m$, $L=10^{19}m$ and $T_{E}=\frac{Z_{E}}{c}s$ for $Z_{E}=3\times 10^{24}m$; with these values $|h|\sim \frac{\varepsilon}{R}\sim 10^{-15}$ which is
within the expected accuracy of PTA \cite{jenet}. Similar results are obtained for other close values of $T_{E}$ }
\label{time2}
\end{figure}

Let us consider the arguments of the trigonometric functions in (3) and define
\be\label{arg}
\bal
\Theta(x,T_E,L,\alpha,\beta,Z_{E},w,\Lambda)\equiv &w(T_E+\frac{L}{c}x-\frac{Z_{E}}{c}-x\frac{L}{c}\cos{\alpha})\\
&+w\sqrt{\frac{\Lambda}{3}}\left(\frac{(\frac{Z_{E}}{c}+
x\frac{L}{c}\cos{\alpha})^2}{2}-\left(T_E+\frac{L}{c}x\right)\left(\frac{Z_{E}}{c}+x\frac{L}{c}\cos{\alpha}\right)\right).
\eal
\ee
Then
\be\label{tres2}
\bal
H(T_E,L,\alpha ,\beta ,Z_{E},w,\varepsilon,\Lambda)=&-\frac{1}{2}\frac{L\varepsilon}{c}\left(\sin^2{\alpha}\cos^2{\beta}+2\sin{\alpha}\sin{\beta}\cos^2{\beta}-\sin^2{\alpha}\sin^2{\beta}\right)\\
&\int_{-1}^{0}dx\frac{1}{(Z_{E}+x L\cos{\alpha})}\left( 1+\sqrt{\frac{\Lambda}{3}}(T_E+\frac{L}{c}x)\right)\left(\cos\Theta +\sin\Theta\right).
\eal
\ee
At this point one should ask whether the observationally preferred exceedingly small value of the 
cosmological constant \cite{cc} affects the timing residuals from a pulsar at all. 
To answer this question we take reasonable 
values of the parameters both for the GW and one pulsar location and plot a snapshot of the resulting timing residuals as a 
function of the angle $\alpha$ for the time of arrival of the wave to the local system, $T_{E}$. The comparison is shown in Figure \ref{time2}.
The figure speaks by itself and it strongly suggests that the angular dependency of the timing 
residual is somehow influenced by the value of the
cosmological constant, in spite of its small value. Another feature that catches the eye immediately is an 
enhancement of the signal for a specific small angle $\alpha$ (corresponding generally to a source of low
galactic latitude, or a pulsar nearly aligned (but not quite as otherwise $E_{ij}\hat n^{i}\hat n^{j}=0$, although the total timing residual is non-vanishing due to the $\mathcal{O}(\Lambda)$ time components for TT waves) with the source. 

\begin{figure}[htb]
\centering
\begin{tabular}{cc}
\includegraphics[scale=0.67]{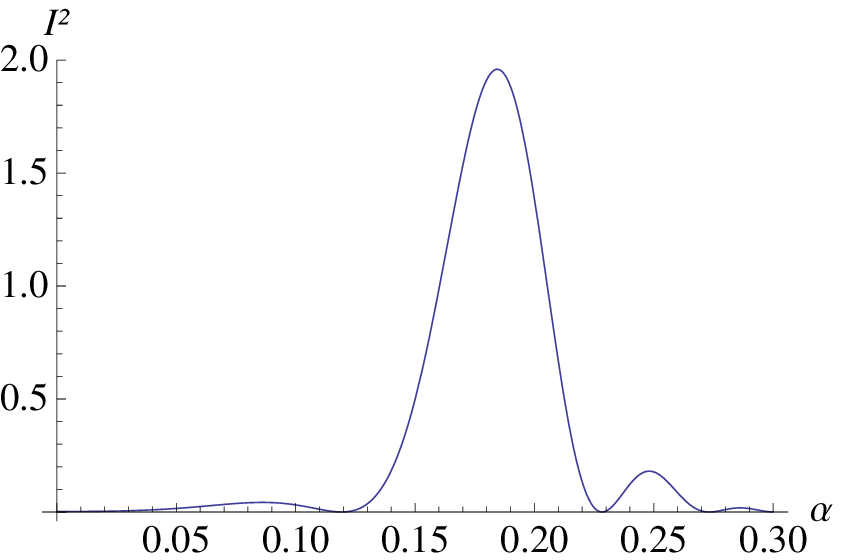} &
\includegraphics[scale=0.67]{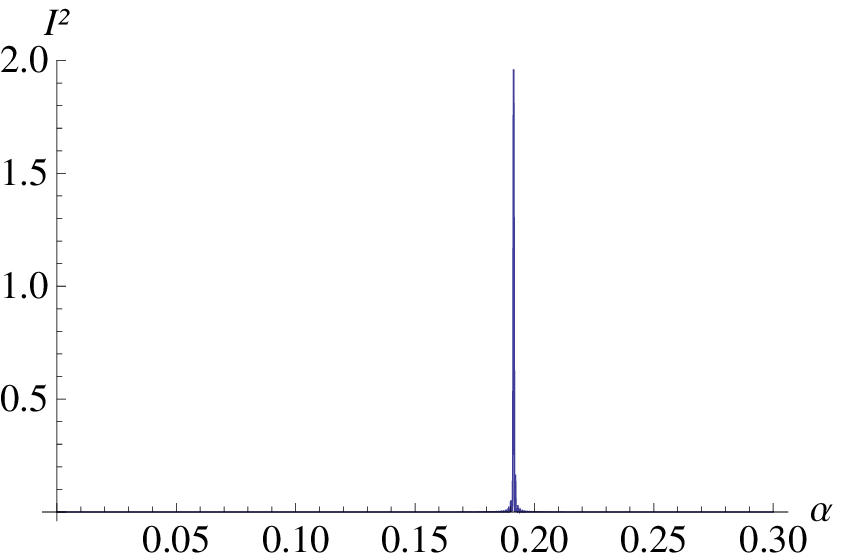}\\
 a) & b)\\
\includegraphics[scale=0.67]{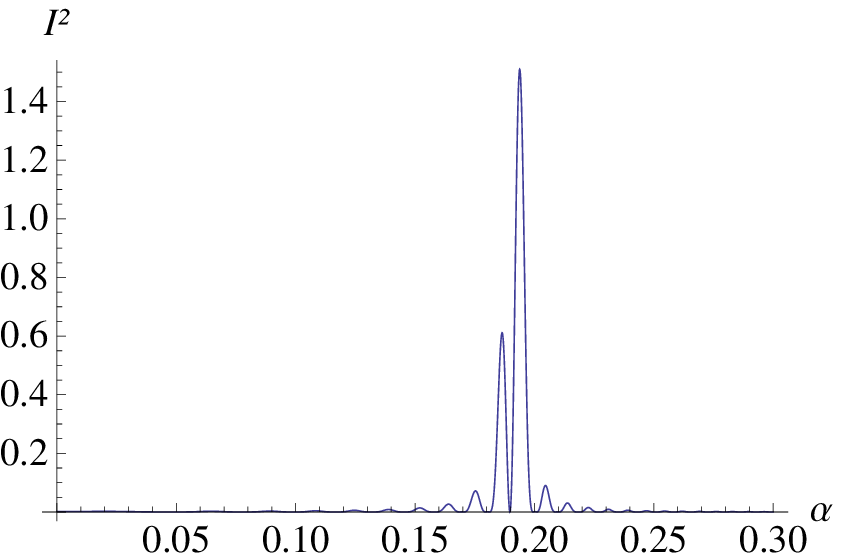} &
\includegraphics[scale=0.67]{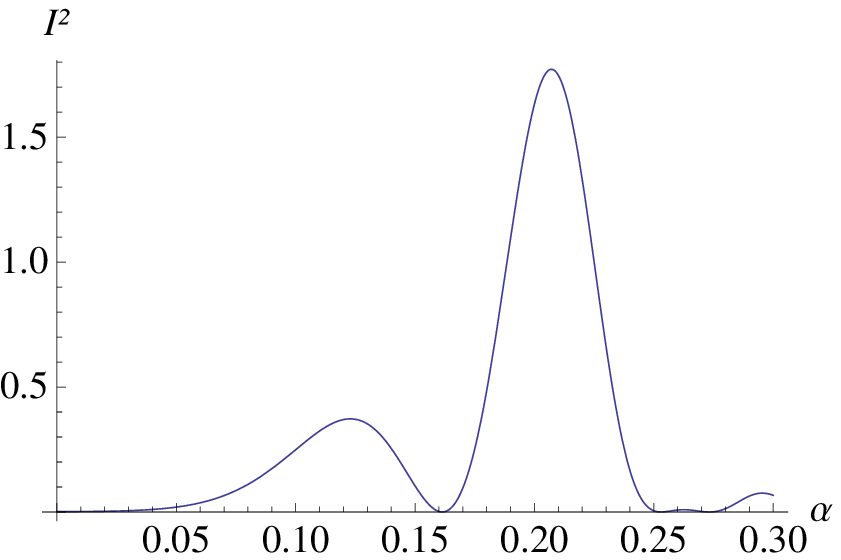}\\
 c)& d)\\
 \includegraphics[scale=0.67]{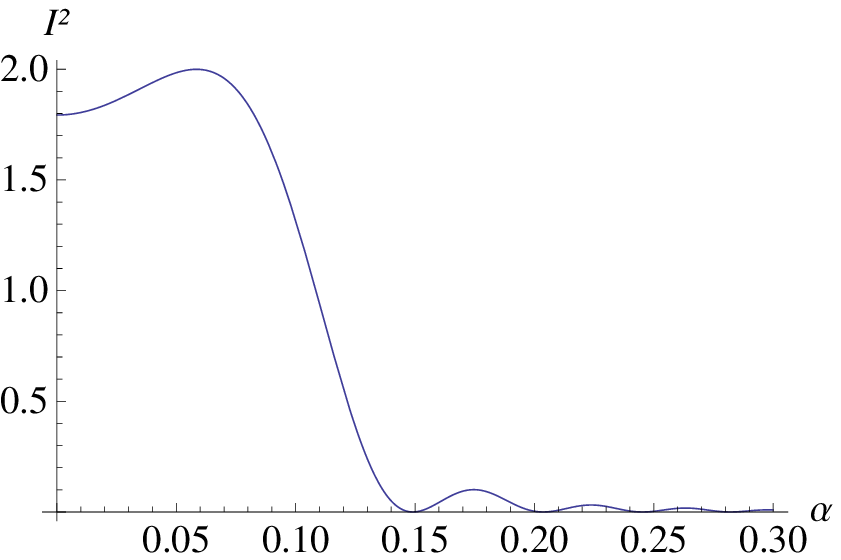} &
 \includegraphics[scale=0.67]{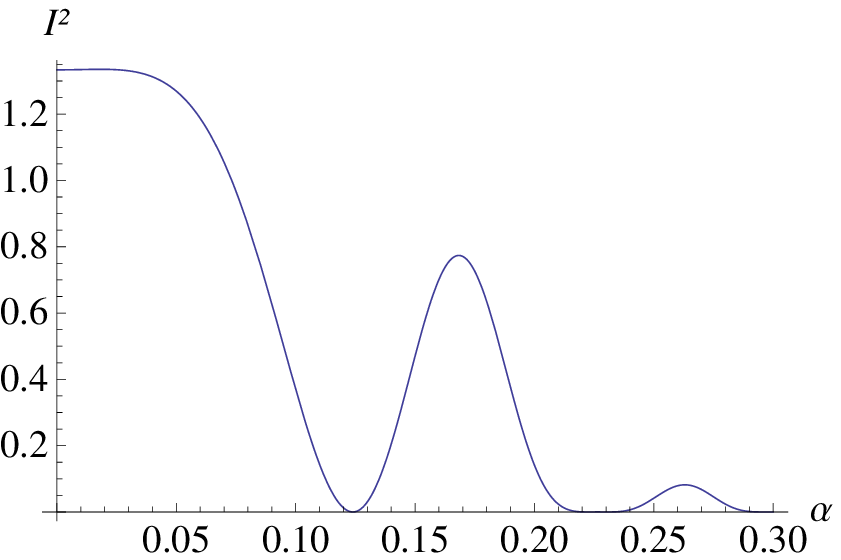}\\
 e)& f)\\
\end{tabular}
\caption{Integral $I^2$ plotted for different values of the parameters involved. 
a) Corresponds to the reasonable values $w=10^{-8}s^{-1}$, $L=10^{19}m$, $Z_{E}=3\times 10^{24}m$, $\Lambda=10^{-35}s^{-2}$ and $T_{E}=10^{16}s$. 
b) Change in pulsar distance to $L=10^{21}m$. 
c) Change in frequency to $w=10^{-7}s^{-1}$. 
d) Change in time to $T_{E}=(10^{16}+10^{8})s$.
e) Change in time and distance to the source to $Z_{E}=3\times 10^{23}m$ and $T_{E}=10^{15}s$. 
f) Change in the cosmological constant to $\Lambda=10^{-36}s^{-2}$.}
\label{time}
\end{figure}

To understand this enhancement let us analyze the behavior of the integral
\be\label{intrig}
I=\int_{-1}^{0}dx(\cos{\Theta}+\sin{\Theta}),
\ee
with $\Theta$ defined in (\ref{arg}) as the prefactors in (\ref{tres2}) are not relevant for the discussion. The result 
can be expressed as a combination of Fresnel functions, and sines and cosines. 
In the limit where $\Lambda\rightarrow 0$ the Fresnel functions go to a constant and the behavior is 
the usual for trigonometric functions. In this respect, the Fresnel functions are responsible for the position 
and magnitude of the enhancement. This is clearly seen when $I^2$ is plotted\footnote{We plot $I^2$ rather than $I$ to deal with a positive quantity.} as a function of the angular 
separation $\alpha$ between the source and the pulsar. $I^2$ always shows a maximum, the position of which 
is quite stable under changes of most of the parameters involved. It turns out to only depend strongly 
on the value of $\Lambda$ and on the distance to the source. It actually depends on the time 
scales involved rather than on the distance to the source but since the time of arrival of 
the wave to the local system is directly related to the distance, the dependency is correlated. 
This is evidenced in Figure \ref{time}
which shows plots of $I^2$ for different values of the frequency, distance to the 
pulsar, distance to the source and cosmological constant. In Figure \ref{time}.a the following reasonable 
values, $Z_{E}=3\times 10^{24}m$, $w=10^{-8}s^{-1}$, $T_{E}=\frac{Z_{E}}{c}s$ and $L=10^{19}m$ are used. 
In b) there is a change in the distance to the pulsar. In c) we change the frequency. In d) we keep the distance 
to the source fixed and use the time at the end of an hypothetical 3 year observation. In e) we change the distance
 to the source one order of magnitude (therefore time also changes). Finally in f) the cosmological constant 
is changed. It is clear that the most dramatic changes occur when either the distance to the source or the value 
of the cosmological constant are modified.

\section{Significance of the timing residuals}
Now we would like to make a more detailed study of this possible signal. For that we use the ATNF pulsar 
catalogue \cite{cata}. 
As it is well known pulsars are remarkably stable clocks whose periods are known to a very high accuracy, 
up to  $10^{-14}s$ in some cases. 
However to achieve this extreme precision requires some hypothesis that are not appropriate for the physical situation
we are considering and we will assume the more modest precision of $\sigma_{t}=9.6\times 10^{-7}\sim10^{-6}s$. This value is obtained by averaging the precision achieved for the best measured pulsars included in the International Pulsar Timing Array Project, Table 1 in \cite{hobbs3}. We are aware that only around forty pulsars are monitored with such accuracy for the time being. Any future improvements in precision and scope would directly translate into an improvement of the results presented in this section.

For each pulsar we have the galactic latitude ($\phi$), the galactic longitude ($\theta$) and the distance ($L$) to the Earth. 
We transform these coordinates to ($\alpha$,$\beta$), where $\alpha$ as already explained is the angular separation 
between 
the line Earth-GW source and the line Earth-pulsar. $\beta$ corresponds to the azimuthal angle of the pulsar 
referred to the plane perpendicular to the line Earth-source.

The statistical significance of the timing residual will be
\be\label{signi}
\sigma=\sqrt{\frac{1}{N_{p}N_{t}}\sum_{i=1}^{N_{p}}\sum_{j=1}^{N_{t}}\left(
\frac{H(T_E^{i,j},L_{i},\alpha_{i} ,\beta_{i} ,Z_{E},w,\varepsilon,\Lambda)}{\sigma_{t}}\right)^2}
\ee
where $\sigma_{t}$ is the accuracy with which we are able to measure the pulsar signal period. 
We take $\sigma_{t}=10^{-6}s$ as mentioned. The index $i$ running from 1 to $N_p$ labels the pulsars
included in the average.

In the statistical average we assume an observation time span of approximately 3 years, 
starting at the time the signal is $10^{16}s$ old 
(time of arrival at our Galaxy). We assume that we perform observations every 11 days. 
That is $N_{t}=101$; $10^{16}s\leq T_{E} \leq 1.00000001\times 10^{16}s$. Since the coalescence times of super massive black hole binaries (SMBHB) can be of the order of $10^{7}s$ \cite{coal} (that is a much shorter time scale than the time of arrival of the perturbation to the local system) it is justified to use $T_{E}=\frac{Z_{E}}{c}$. Form Figure \ref{time}d one can also see that the position of the enhancement is not significantly altered in the time span of observation. While we are aware that 3 years is a short time (most 
studies consider observational periods from 5 to 10 years) we do not intend to present here but
a proof of principle and we prefer to consider a short period for our numerical analysis. Longer periods
of observation will of course reinforce the signal.

We turn to the angular dependence of the significance. 
In the following $\sigma(\alpha)$ is plotted keeping $\alpha$ as a free parameter (note that it is not summed up), that is, using a set of 5 fixed pulsars supposed to be exactly at the same angular separation\footnote{The chosen pulsars belong to a globular cluster that in principle can not be currently timed with the assumed accuracy due to the internal accelerations within the cluster, in this respect this is still a theoretical exercise. We thank the referee for pointing this to us.} from a source the position of which we vary $0\leq\alpha\leq \pi$ (the result again is symmetrical for $\pi\leq\alpha\leq 2\pi$). This could be done for any set of five pulsars, since, as shown in the previous section, the position of the peak does not depend on the values $L_{i}$ and $\beta_{i}$. However, we used the following set of real pulsars which are all close to each other at a distance $L\sim 10^{20}m$. 
\begin{table}[htb]
\centering
\begin{tabular}{|c|}
\hline
Pulsars from the ATNF Catalogue\\
\hline
  J0024-7204E	\\
  J0024-7204D	\\
  J0024-7204M	\\
  J0024-7204G	\\
  J0024-7204I	\\
\hline
\end{tabular}
\caption{List of pulsars whose $L_{i}$ and $\beta_{i}$ we used to calculate $\sigma(\alpha)$ for an hypothetical source at angular separation $\alpha$.}
\label{taula}
\end{table}
It must be borne in mind that although there are over 600 pulsars, making it easy to find clusters with a similar $\alpha$ (albeit possibly with very different
values of $L$ and $\beta$), the precision with which they are timed can vary widely. The overall magnitude of the significance depends directly on the precision of the measured period as well as on the amplitude of the wave. The results presented in the following may not be entirely realistic due to the uncertainties in these values but the general features of the analysis would remain unchanged if more realistic values (eventually available) are used.
\be\label{sigal}
\sigma(\alpha)=\sqrt{\frac{1}{5\cdot 101}\sum_{i=1}^{5}\sum_{j=1}^{101}\left(
\frac{H(T_E^{i,j},L_{i},\alpha ,\beta_{i},3\times10^{24},10^{-8},1.2\times 10^{9},10^{-35})}{10^{-6}}\right)^2}.
\ee
Length units are given in meters, frequencies in $s^{-1}$.
\begin{figure}[htb]
\centering
\includegraphics[scale=1.13]{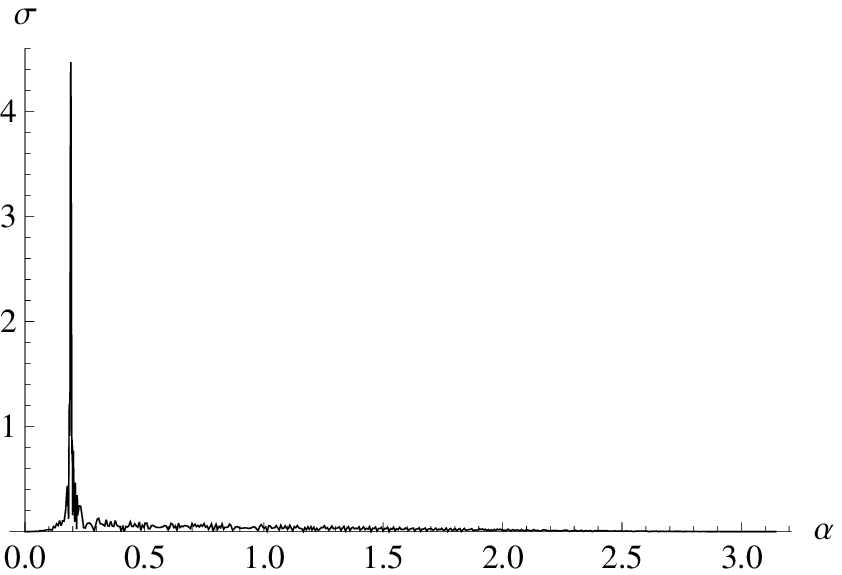}\\
\includegraphics[scale=1.13]{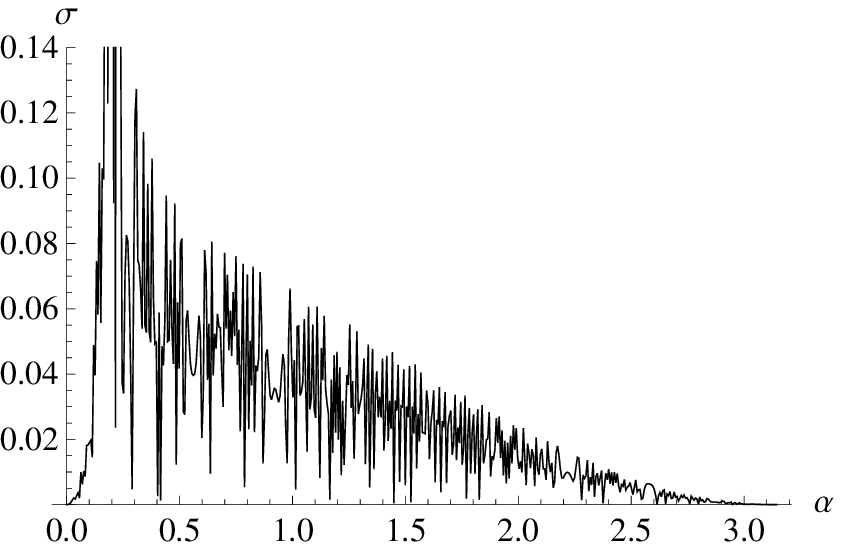}\\
\includegraphics[scale=1.13]{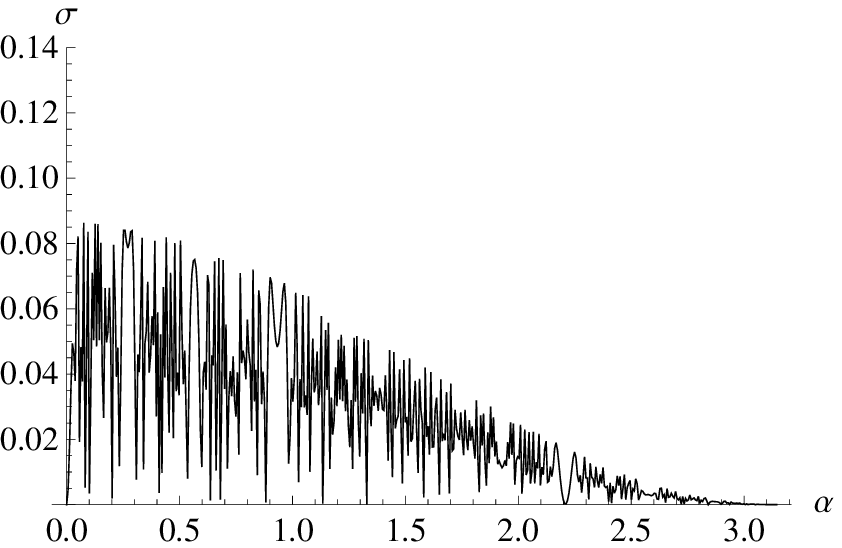}
\caption{$\sigma(\alpha)$ for $\Lambda=10^{-35}s^{-2}$ (Top). Zoom on the lower values for $\Lambda=10^{-35}s^{-2}$ (middle), and comparison to $\Lambda=0$ (bottom).}
\label{sigdal2}
\end{figure}
We observe a huge peak at $\alpha\sim 0.19$rad (see Figure \ref{sigdal2}). If a source is located 
at such angular separation\footnote{The angular position of the enhancement corresponds approximately to a stationary point of the phase of the wave, i.e. the path along which the phase of the gravitational wave is practically constant, making the integrated timing residual maximal. If one considers the wave front of a Minkowskian spherical wave, there is no path parameterized as the argument in (4), corresponding to a straight line in space, with a constant phase (other that paths pointing to the source, for which the amplitude is zero). Instead, if one considers a wave with a frequency and wave vector depending on the space-time coordinates, it is in principle possible to find a path where the changes in frequency and wave vector compensate the change in phase, making the overall phase constant along that path. It can be seen by deriving the phase of the trigonometric function in $h_{ij}$ with respect to $x$ and imposing the stationary phase condition that one gets a solution for $\alpha\neq 0$ that depends very weakly on $x$ itself.} from the average angular position of the five pulsars chosen for observation, 
the significance could be boosted some 50 times. Let us compare it to the same calculation 
taking $\Lambda=0$ and redshifted frequency $w_{eff}=\frac{w}{(1+z)}$; $z\sim 0.008$, which is the corresponding 
redshift for an object $10^{24}m$ away calculated using both matter and energy densities. No peak is observed.

Now we take a list of observed pulsars well distributed in the galaxy. The angles $(\alpha,\beta)$ are calculated 
for all of them considering two hypothetical sources of GW. One located 
at galactic coordinates $\theta_{S1}=300^{\circ}$, $\phi_{S1}=-35^{\circ}$ and another located 
at $\theta_{S2}=4^{\circ}$, $\phi_{S2}=10^{\circ}$. We order the pulsars from the lowest $\alpha$ to the largest. We 
group them in sets of five pulsars. We consider 27 sets of 5 pulsars; that is a list of 135 pulsars. 
For each set we calculate the significance
\be
\sigma_{k}=\sqrt{\frac{1}{5\cdot 101}\sum_{i=1}^{5_{k}}\sum_{j=1}^{101}\left(
\frac{H(T_E^{i,j},L_{i},\alpha_{i} ,\beta_{i},3\times10^{24},10^{-8},1.2\times 10^{9},10^{-35})}{10^{-6}}\right)^2}
\ee
and plot it as a function of the average angle of the set, $\bar{\alpha}_{k}=\sum_{i=1}^{5_{k}}\frac{\alpha_{i}}{5}$ with $1\leq k \leq 27$. Note this is different from (\ref{sigal}); here we choose two hypothetical fixed sources 
and a long list of pulsars grouped by their angular separation $\alpha$ to these sources. 
This could be a realistic calculation once real sources are considered.

\begin{figure}
\centering
\includegraphics[scale=0.7]{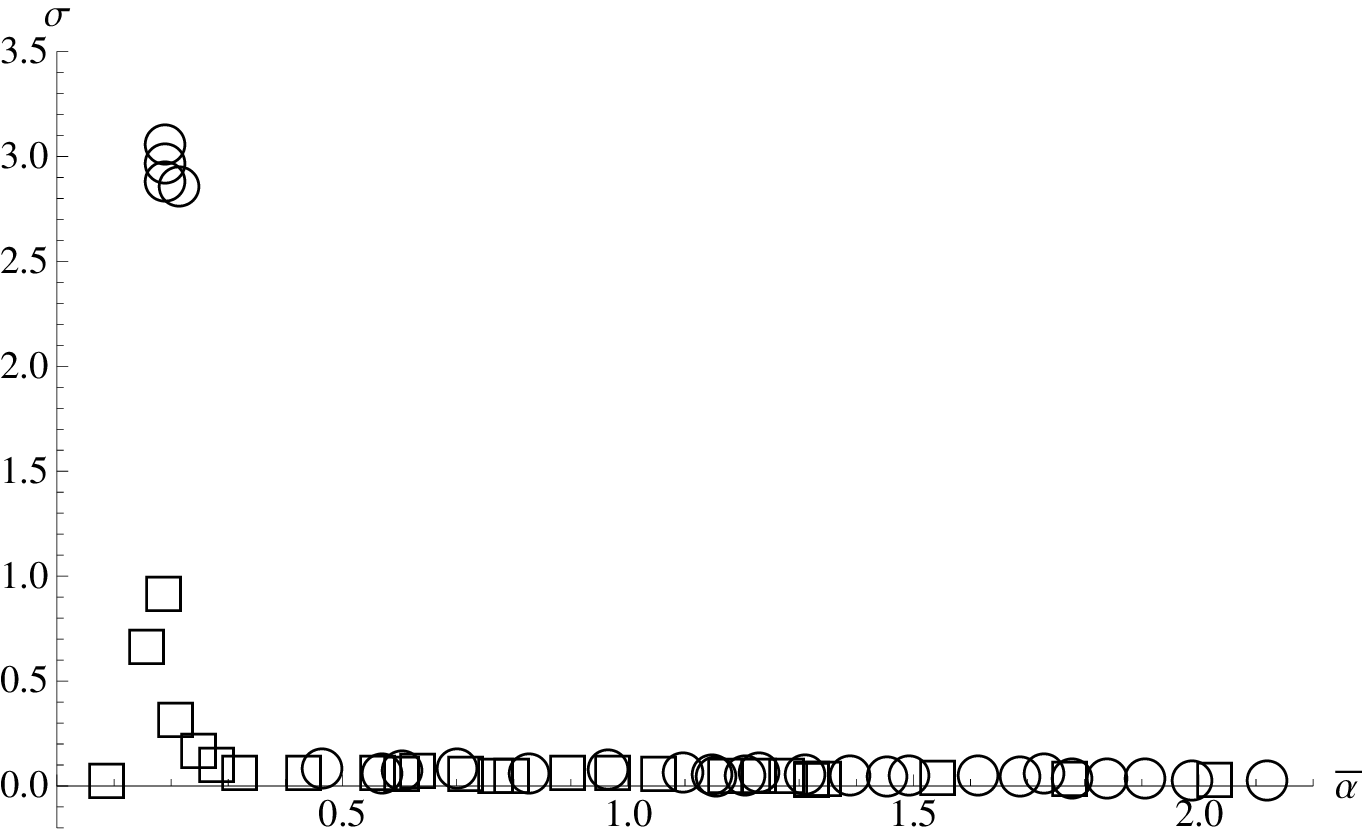}\\
\includegraphics[scale=0.7]{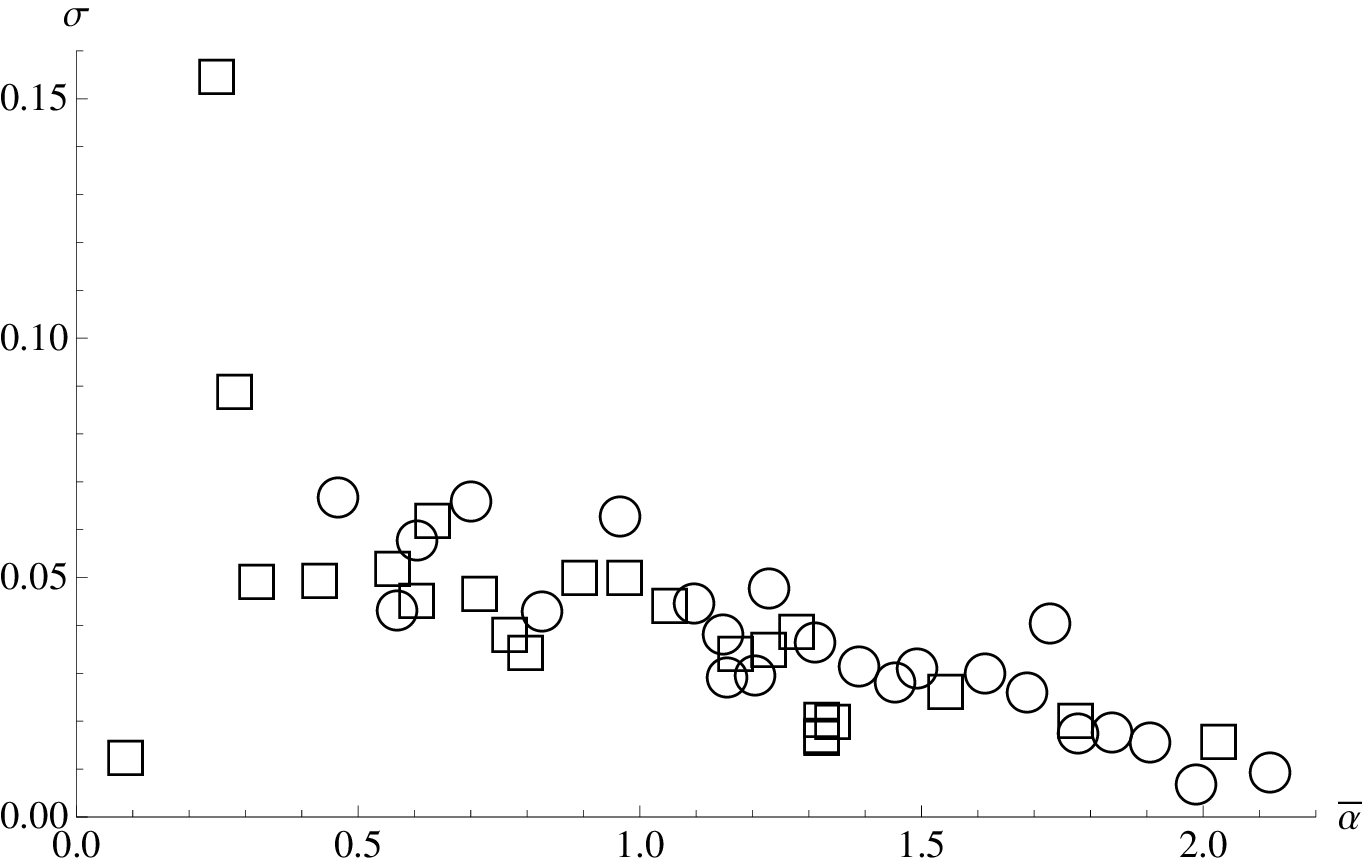}\\
\includegraphics[scale=0.7]{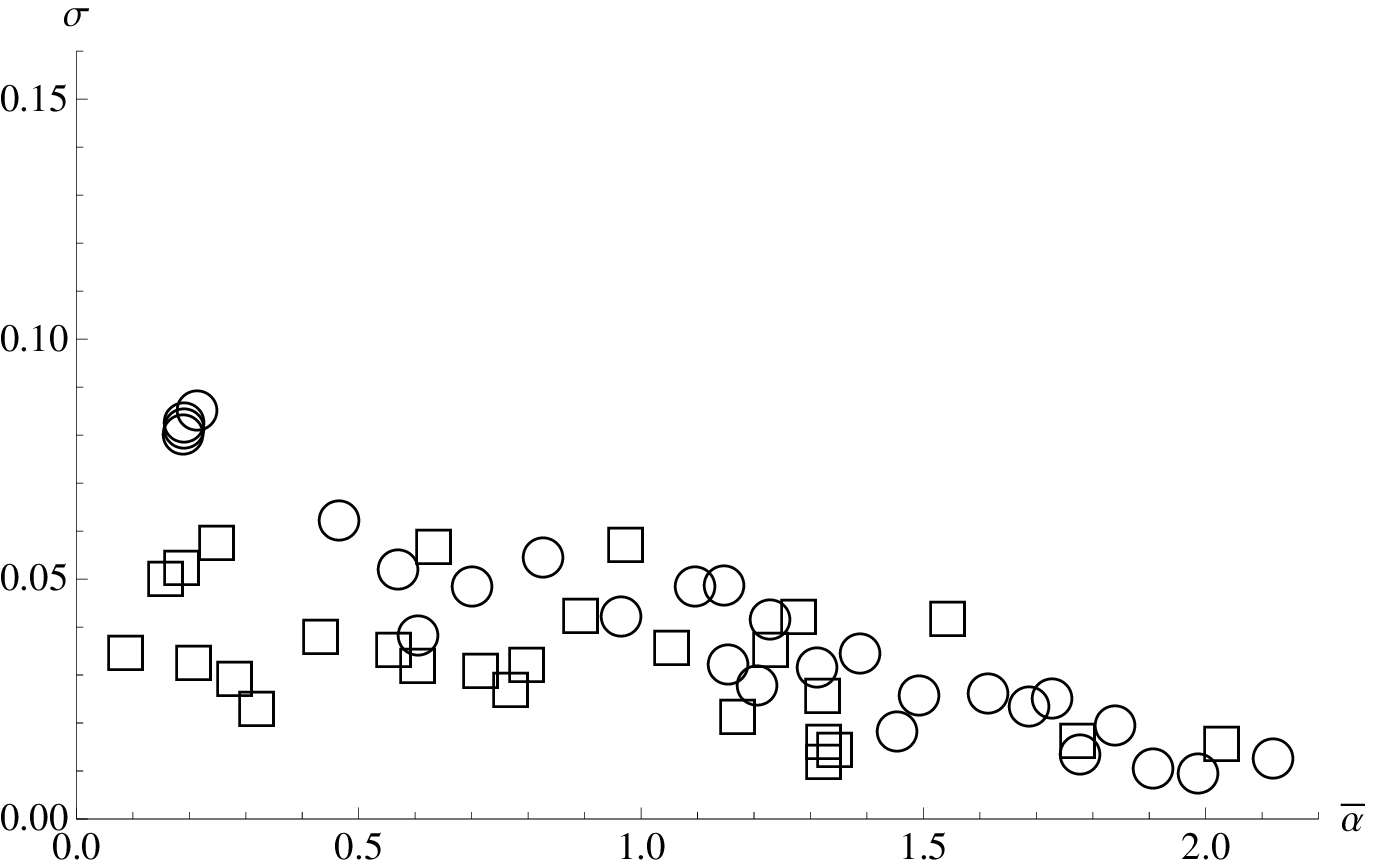}
\caption{Plot of $\sigma_{k}(\bar \alpha_{k})$, $k=1,27$. $\Lambda=10^{-35}s^{-2}$. Circles correspond to Source 
1 and squares to Source 2. Full range is showed on top, zoom on the lower values for $\Lambda=10^{-35}s^{-2}$ and 
comparison to $\Lambda=0$ show on middle and bottom respectively.}
\label{picp}
\end{figure}

The results obtained are plotted in Figure \ref{picp}. In both cases a very noticeable peak 
is observed at the expected angle.

The reason why the peak for Source 2 is lower than the peak for Source 1 is that Source 2 is located close but not at the precise angular separation of a real cluster of pulsars. 
This is meant to illustrate that even in that case a significant enhancement of the signal can be achieved. 

\begin{figure}[htb]
\centering
\includegraphics[scale=0.71]{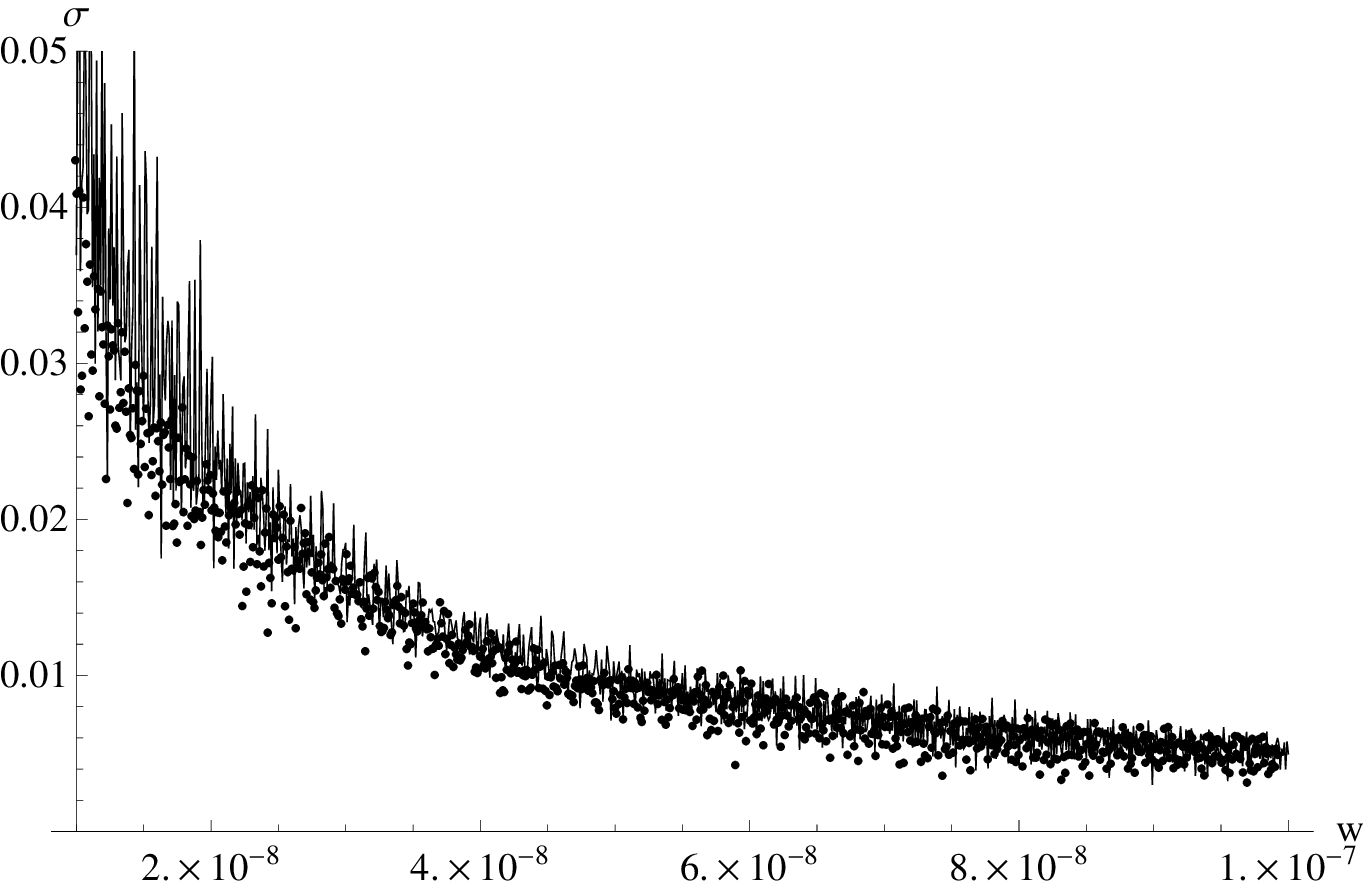}\\
\includegraphics[scale=0.7]{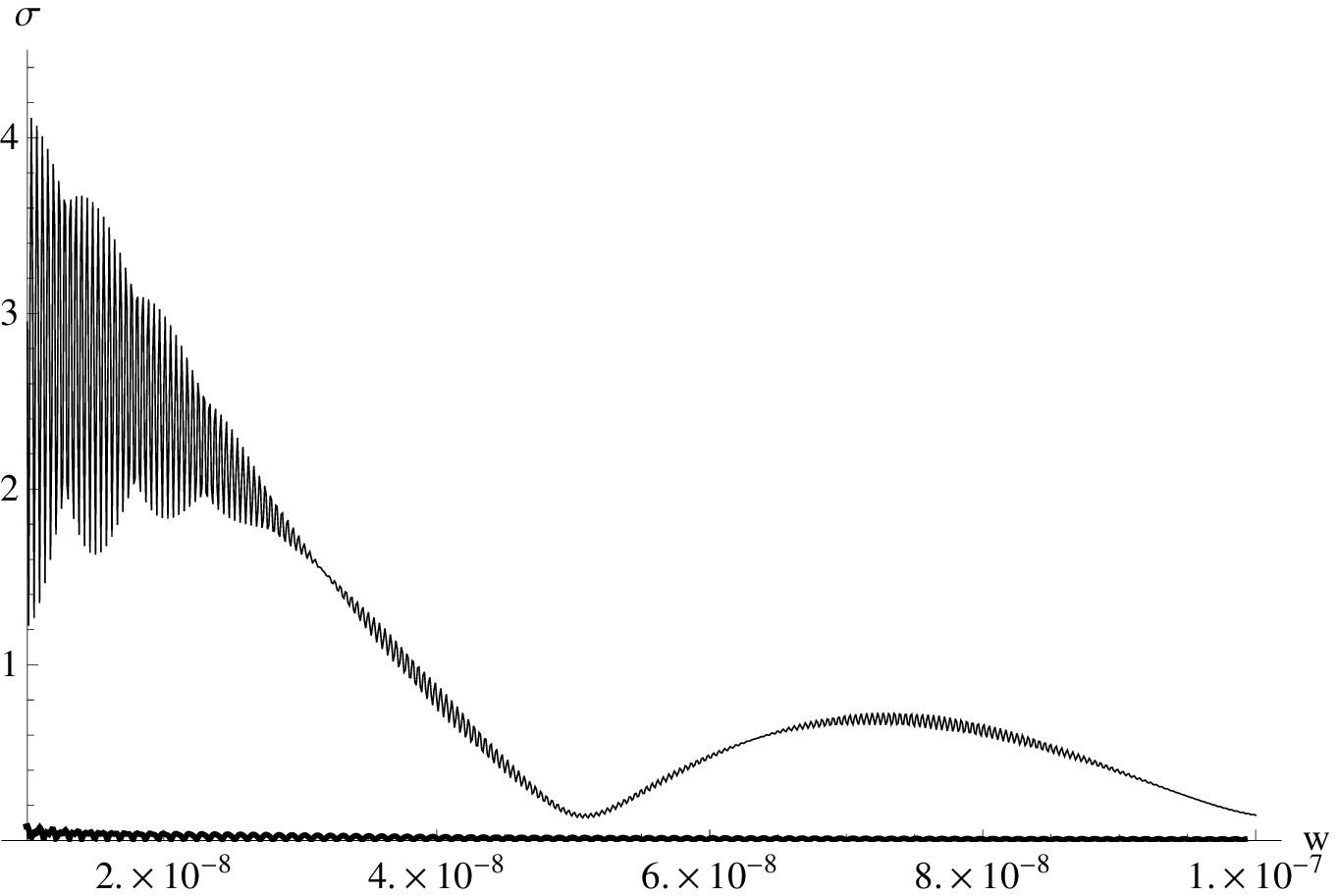}\\
\includegraphics[scale=0.7]{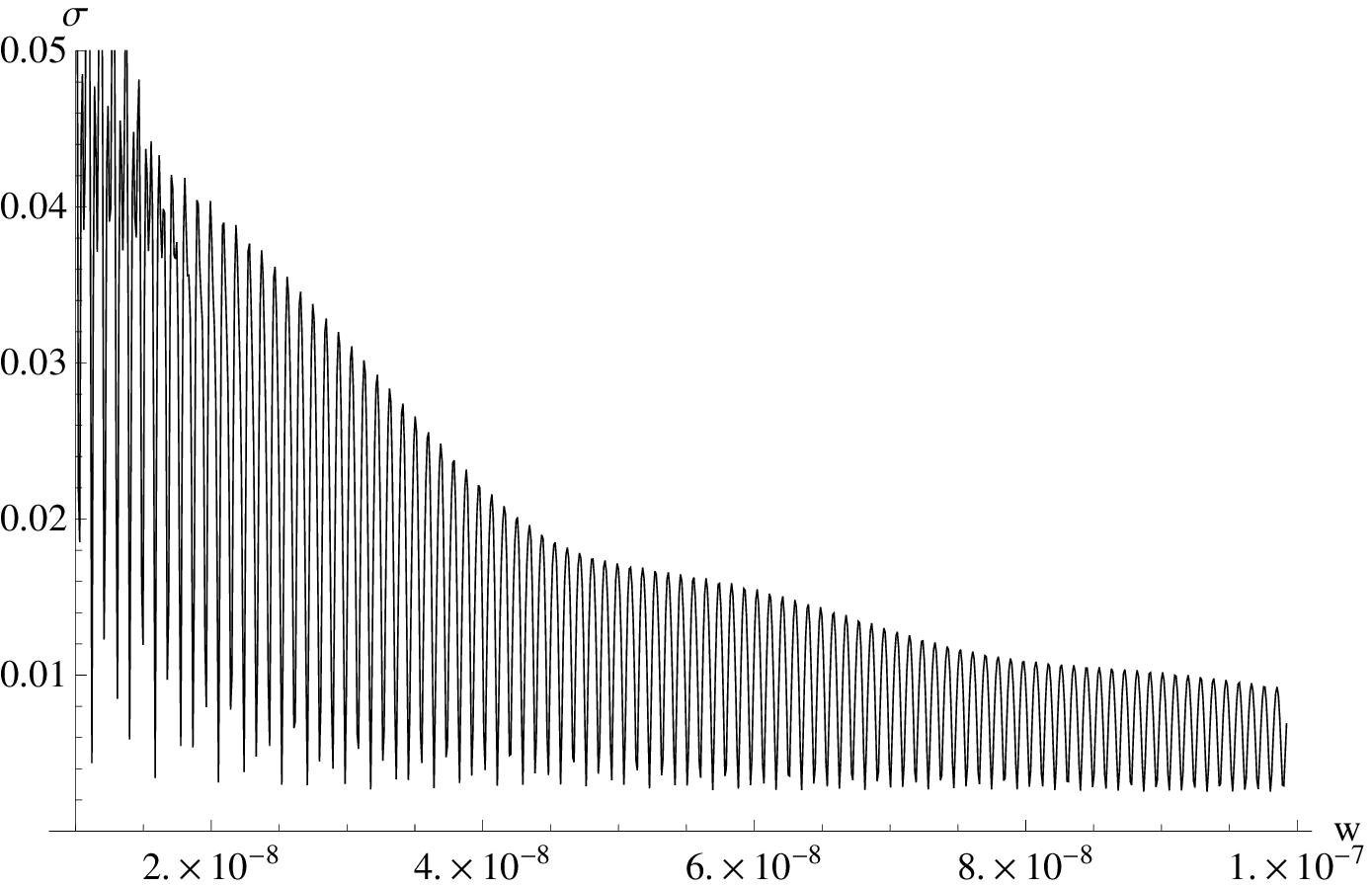}
\caption{$\sigma(w)$ for 15 pulsars away from the peak angular region for Source 1 (top), the solid line corresponds to $\Lambda=10^{-35}s^{-2}$ and dots correspond to $\Lambda=0$. $\sigma(w)$ for 15 pulsars at the peak angular region for the same source (middle). Solid line corresponds to $\Lambda=10^{-35}s^{-2}$ and the data close to the horizontal axis 
corresponds to $\Lambda=0$. Zoom on the $\Lambda=0$ case (bottom). }
\label{power}
\end{figure}

Finally, the dependency of $\sigma$ on the frequency
\be
\sigma(w)=\sqrt{\frac{1}{N_{p}\cdot 101}\sum_{i=1}^{N_{p}}\sum_{j=1}^{101}\left(
\frac{H(T_E^{i,j},L_{i},\alpha_{i} ,\beta_{i},3\times10^{24},w,1.2\times 10^{9},10^{-35})}{10^{-6}}\right)^2},
\ee
has also been investigated. Some of our preliminary checks indicated that no differences at all were observed in the power spectrum when the value of $\Lambda$ was changed and that, as expected \cite{hobbs2,hobbs3,jenet}, the signal follows a power law $\sigma\sim \frac{1}{w}$. However, let us take a closer look at the dependency on the frequency for a short list of pulsars located at the right angular separation to observe the peak. We have already seen the significance grows notoriously in this angular region. Figure \ref{power} (middle) shows the frequency dependence of the signal for fifteen pulsars at the right spot with respect to Source 1. As it can be clearly seen, the signal significance grows enormously again for $\Lambda=10^{-35}s^{-2}$ and apparently does not follow a power law. For the same short list of pulsars and for $\Lambda=0$ the signal falls back to smaller values and its envelope shapes towards a power law. In Figure \ref{power} (top) we also present the same plot for fifteen pulsars located at an angular separation of around $\alpha\sim 1.1$rad, that is away from the peak. In this case we see no differences between the different values of the cosmological constant as well as a clear power law behavior. The magnitude of the signal is compatible with that of the fifteen pulsars at the peak separation when $\Lambda=0$.

\section{Measuring the cosmological constant}
We have seen in the previous that there is an enhancement in the timing residual for a particular 
value of the angle $\alpha$ when GW propagating in de Sitter space-time are measured. 
Among all the dependencies, and when the distance to the source is well known, the most relevant appears to
 be the one related to the value of the 
cosmological constant $\Lambda$. The position of the peak depends strongly on the value of $\Lambda$. 
It moves towards the central values of the angle for larger values of the cosmological constant. 

\begin{figure}[htb]
\centering
\includegraphics[scale=0.9]{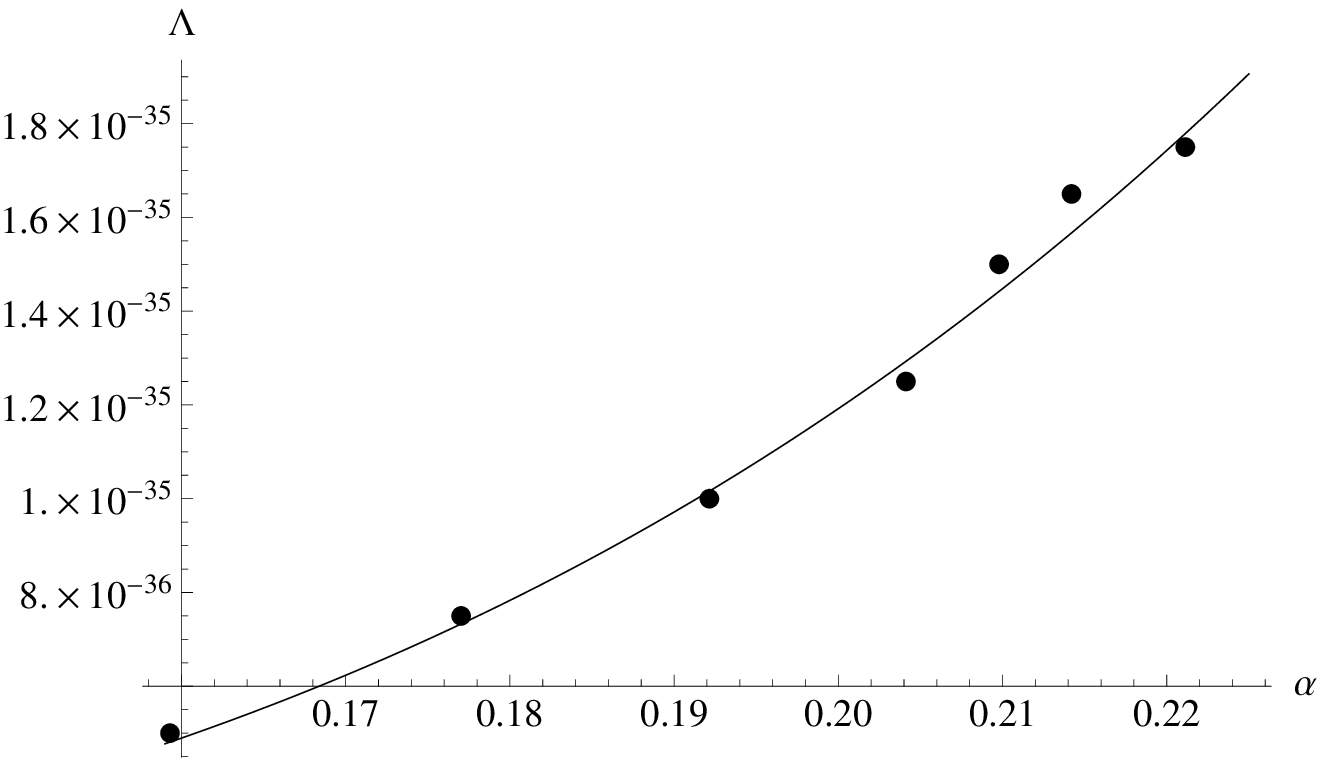}
\caption{$\Lambda(\alpha)$ obtained numerically from the positions of the peaks in the $\sigma(\alpha)$ plots for different values of the cosmological constant (dots) and obtained analytically from an approximation of the Fresnel functions involved in the timing residual (line).}
\label{lamdal}
\end{figure}

The values of $\Lambda$ as a function of the position at which the peak would be found 
are plotted in Figure \ref{lamdal} (dots) using the positions found in the plots for $\sigma(\alpha)$ (\ref{sigal}) 
for different values of the cosmological constant. This calculation was carried out using two independent numerical
 methods in order to make sure that one is free of numerical instabilities (this is a necessary precaution as
large numbers are involved). 

We argued in Section 2 that the position of the 
peak is determined by the Fresnel functions one obtains when calculating the timing residuals. Indeed the integral $I$ in
(\ref{intrig}), which captures the crucial effect, gives a prefactor times a combination of Fresnel functions times a combination of ordinary trigonometric functions. The latter are featureless; however the prefactor becomes 
quite large for a specific 
value of the parameters involved. This particular value renders the Fresnel function close to zero and the product is
a number close to 2. Away from this point the net result is small. 

Using the series expansion of the Fresnel functions at first order we are able to obtain an 
approximate analytical expression for the relation $\Lambda(\alpha)$; that is for the value of the cosmological constant
that (all other parameters being fixed) gives a strong enhancement of the significance $\sigma$ at a given angle $\alpha$
\be\label{lam}
\Lambda(\alpha)=\frac{12 c^2 \sin ^4\left(\frac{\alpha }{2}\right)}{\left((c T_{E}-Z_{E})\cos \alpha  +Z_{E}\right)^2} \simeq \frac{12 c^2 \sin ^4\left(\frac{\alpha }{2}\right)}{Z_E^2},
\ee
which is also shown in Figure \ref{lamdal} (line). We have used the fact that, taking into account the duration 
of a black-hole merger, $cT_E\simeq Z_E$.
Eq. (\ref{lam}) is a clear prediction that could be eventually tested. In fact, this effect could also
facilitate enormously the detection
of GW coming from massive binary black holes by carefully selecting and binning groups of pulsars, although the possibility
of measuring $\Lambda$ locally certainly looks to us more exciting. 

Throughout this work we have considered only the effect of $\Lambda$ on GW and the way they affect
pulsar timing residuals and we have neglected the effect of matter or matter density. In fact, the main 
effect of the latter would be through the familiar redshift in the frequency of GW. The overall frequency value does not play a crucial role in the previous
discussion provided that is low enough to be detectable in PTA. It is its dependency on the space-time coordinates thats brings about new effects. It is probably useful to remind the reader once more that $\Lambda$ is assumed to be an intrinsic property of 
space-time, present to all scales, except close to the Galaxy. It would be easy to implement more realistic models in our 
study, if reasonably well-defined ones were available. In fact, these uncertainties constitute 
strong reasons to try to measure
$\Lambda$ locally.

\section{Summary}
The purpose of this work was to investigate the local effects of the cosmological constant 
for the detection of gravitational waves in PTA. The gravitational wave function is usually modeled 
as a massless, either plane or spherical, wave traveling in flat space-time. 
The expansion of the universe is accounted for by including a redshift in the frequency. 
Major problems are related to modeling the source and assessing the strain of the amplitudes of the waves. 
Here we obviate these by just assuming a spherical wave and focus in the fact that the waves propagate 
in a de Sitter space-time rather 
than in flat space-time. 

We use a wave solution previously derived in FRW coordinates, which we expect to be more realistic 
than the redshifted usual waves. With this, we calculate the timing residuals induced in the signal 
of known pulsars in our Galaxy, predicting a particular value of the angle subtended between the source and the
pulsar where an enhanced significance of the timing residual is observed. We argue that the position of this 
peak depends strongly on the value of the cosmological constant. This peak is absent when the calculations are 
carried out with usual, Minkowski solution, redshifted waves. We propose two hypothetical sources at 
two distinct positions 
for which we calculate the timing residuals significance using a real set of pulsars. 
The peak is observed at the predicted angular position. Finally we obtain the angular dependency 
of the value of the cosmological constant using the position of the peak for different 
values of $\Lambda$ and analytically from the Fresnel functions involved in the calculation. This method 
could represent an independent way to determine the value of the cosmological constant.

We should end up with a disclaimer. The results presented in this article are by all means preliminary. 
We have addressed the somewhat academic study of an isolated point-like source of GW and proceeded to 
analyze its influence on PTA. In fact, we should expect a complete background of sources. Preliminary 
studies indicate that single source detection may be feasible, but at values of the red shift larger
that those considered here \cite{hobbs3}. Of course these studies do not
consider the effects of $\Lambda$ discussed in this paper. In addition a full error analysis including all unknowns
in the appropriate covariance matrix should be performed before drawing conclusions on the statistical
significance of the effect. Some of these studies are only feasible by the PTA collaborations themselves
whose interest on the effect presented here we hope to have aroused.

\section*{Acknowledgements}
We acknowledge the financial support from research Grants FPA2007-20807, SGR2009SGR502. This research is supported by the Consolider CPAN project. We thank Ll. Garrido, J. Bernabeu, J.M. Paredes, J. Salvad\'o and J. Mold\'on for useful discussions on the subject. We also thank L. Verde for reading the manuscript and A. Sintes for extended discussions and for pointing to us several references.

\end{document}